\documentclass[aps,prd,twocolumn,superscriptaddress,preprintnumbers,showpacs]{revtex4-2}
\usepackage{amsmath,amsfonts,amssymb,graphicx,color,times,bbm,psfrag,dsfont,
slashed,bigints,bm, verbatim}
\usepackage[colorlinks=true,citecolor=blue]{hyperref}
\usepackage[normalem]{ulem}

\newcommand{\be}{\begin{equation}}
\newcommand{\ee}{\end{equation}}
\newcommand{\ba}{\begin{eqnarray}}
\newcommand{\ea}{\end{eqnarray}}

\newcommand{\mc}[1]{\textcolor{black}{#1}}
\newcommand{\mm}[1]{\textcolor{black}{#1}}
\newcommand{\dg}[1]{\textcolor{black}{#1}}

\usepackage[normalem]{ulem}
\usepackage{xcolor}

\begin{document}

\title{Hawking temperature and phonon emission in acoustic holes}

\author{Massimo Mannarelli}
\affiliation{INFN, Laboratori Nazionali del Gran Sasso, Via G. Acitelli,
22, I-67100 Assergi (AQ), Italy}
\author{Dario Grasso}
\affiliation{INFN Sezione di Pisa,
Polo Fibonacci, Largo B. Pontecorvo 3, 56127 Pisa, Italy}
\affiliation{Dipartimento di Fisica, Universit\`a di Pisa,
Polo Fibonacci, Largo B. Pontecorvo 3}
\author{Silvia Trabucco}
\affiliation{Dipartimento di Fisica, Universit\`a di Pisa,
Polo Fibonacci, Largo B. Pontecorvo 3}
\affiliation{INFN Sezione di Pisa,
Polo Fibonacci, Largo B. Pontecorvo 3, 56127 Pisa, Italy}
\author{Maria Luisa Chiofalo}
\affiliation{Dipartimento di Fisica, Universit\`a di Pisa,
Polo Fibonacci, Largo B. Pontecorvo 3}
\affiliation{INFN Sezione di Pisa,
Polo Fibonacci, Largo B. Pontecorvo 3, 56127 Pisa, Italy}

\begin{abstract}
Acoustic holes  are the hydrodynamic analogue of standard black holes. Featuring an acoustic horizon, these systems spontaneously emit phonons at the Hawking temperature. We derive the Hawking temperature of the  acoustic horizon by fully exploiting the analogy between black and acoustic holes within a covariant kinetic theory approach.  After deriving the phonon distribution function from the covariant kinetic equations,
 we reproduce the expression of the Hawking temperature  by equating the entropy and energy losses of the acoustic horizon and the entropy and energy gains of the spontaneously emitted phonons.  
Differently from previous calculations we do not need a microscopical treatment of normal modes propagation.
Our approach opens a different perspective on the meaning of Hawking temperature and its connection with entropy, which may allow an easier study of non stationary horizons beyond thermodynamic equilibrium, including dissipative effects.
\end{abstract}
\maketitle
\section{Introduction}
There exists a remarkable analogy between transonic flow in hydrodynamics and black hole (BH) physics~\cite{Unruh:1980cg}, which allows to define  in both contexts  
an event horizon.
In acoustic holes (AHs), the analog of BHs, long-wavelength sound waves cannot move against the fluid flow beyond the surface (the acoustic horizon) where that becomes supersonic. Intriguingly, in both cases 
one can define the Hawking temperature characterizing the spectrum of the particles spontaneously emitted close to the horizon~\cite{Unruh:1980cg, Brout:1995rd}. 
As photons are spontaneously emitted at the BH horizon, AHs can be viewed as a thermal reservoir at fixed Hawking temperature, $T_H$, continuously emitting phonons, corresponding to long-wavelength sonic vibrations~\cite{Brout:1995rd, Corley:1997pr,Saida:1999ap, Himemoto:2000zt,Unruh:2004zk,Barcelo:2005fc, Balbinot:2006ua}. The first expression of the Hawking temperature of an AH was derived by Unruh 
\cite{Unruh:1980cg}: he presented an expression  of $T_H$ as a function of the gradient of the velocity field for a transonic flow in a background with spatially uniform sound speed. Afterwards, this expression was  generalized to 
\be\label{eq:hawking}
T_H = \frac{1}{2 \pi}\left. \frac{\partial |c_s-v|}{\partial n}\right\vert_H\,,
\ee
where $v$ and $c_s$ are the space dependent fluid velocity and sound speed, respectively, and the normal derivative is evaluated at the event horizon, as discussed for instance in~ \cite{Barcelo:2005fc}. 
Remarkably, both numerical \cite{Carusotto} and laboratory \cite{Steinhauer,Steinhauer1} experiments recently verified that phonons spontaneously emitted by an acoustic horizon in Bose-Einstein condensates have a thermal spectrum with temperature $T_H$. 

The analogy between  hydrodynamics  and gravity resides on the fact that the low-energy phonon action can be written as that of a scalar boson propagating in an effective acoustic metric \cite{Unruh:1980cg,Stone:1999gi, Volovik:2000ua}, 
determined by the hydrodynamic flow. 
For both black and acoustic horizons $T_H$ corresponds to the surface gravity $\kappa$. Then, this  analogy  between quasiparticle propagation and gravitational phenomena has been exploited in a number of works using different media~\cite{Barcelo:2005fc}.

Though these analogies are quite rich, it is useful reminding their limitations. Not all aspects of  gravity can be mimicked by an analog model: the matter stress-energy tensor does not obey the Einstein's equations, therefore it does not exist a link between matter energy density and the analog metric. 
%In hydrodynamics the phonons are long-wavelength quasiparticles, therefore their effective field theory has a natural cutoff length scale~\cite{Jacobson:1991gr, Unruh:1994je}. 
Nevertheless, analogies can inspire unexplored perspectives and sprout new ideas.  
Here, we establish that the intimate relation proven for BHs between the Hawking emission and the  horizon area \cite{bardeen1973} holds as well for acoustic horizons. This provides further support to the conjecture that, analogously to BHs, the entropy of AHs is proportional to their horizon area as expected if it arises by information loss~\cite{Bekenstein:1973ur} or by the entanglement of quantum fluctuations on the two sides of the horizon~\cite{Jacobson:1994iw}.  Our derivation fully exploits the analogy between hydrodynamic flow and gravity yielding a  covariant expression of the phonon distribution function 
from  the  appropriate kinetic equations~\cite{LINDQUIST1966487,stewart1969lecture}, formally expressing 
the concept that spontaneously emitted phonons can be viewed as bosons in Unruh's acoustic metric. The usual expression of the Hawking temperature thus pops out as a
simple and straightforward consequence of the entropy area law which, to our knowledge, remained unnoticed so far. 
Likewise, we establish  a clear link between the entropy and the energy variations of the fluid and the phonon entropy and energy densities. Our results are analytically derived in the limit of vanishing bulk temperature: in other words, the only excitations of the fluid are generated by  the acoustic horizon, neglecting any thermal noise.

This paper is organized as follows. In Sec.~\ref{sec:theoretical_setup} we present the general theoretical setup to describe AHs by a kinetic theory approach. In Sec.~\ref{sec:spherical} we discuss the spherical AHs; although unphysical this case displays a manifest analogy with standard BHs. A more realistic geometry is discussed in Sec.~\ref{sec:realistic}. We draw our conclusions in Sec.~\ref{sec:conclusions}. We include two Appendixes containing details of the calculations. In particular in Appendix \ref{sec:append1} we present a few technical details regarding the dimensional reduction close to the acoustic horizon, while in Appendix \ref{sec:append2} we report  the  evaluation of selected thermodynamic quantities.
% - - -- - - - - -- - 
\section{Theoretical setup}
\label{sec:theoretical_setup}
% - - -- - - - - -- - 

The effective low-energy Lagrangian of a fluid  depends on the equation of state and on the space symmetries of the system, as discussed 
in~\cite{Leutwyler:1996er,SCHAKEL_1996,  Son:2002zn, 2008PhRvD..77j3014M}.
 \begin{comment}In the following, in order to be definite \end{comment} 
 In this work, we use the Minkowski metric $\eta_{\mu\nu} = {\rm diag}(1,-1,-1,-1)$ and natural units  $c= \hbar =k_B=1$.
For simplicity, we consider a barotropic equation of state  $P \equiv P (\mu)$,
where $P$ is the pressure and $\mu$ is the  chemical potential. Assuming irrotational flow, the background velocity  configuration can be described by a scalar field
 $\varphi$, and  defining
$D_\rho  \varphi \equiv \partial_\rho \varphi -  \delta_{\rho0} \mu$ with $\delta$ the Kronecker's symbol, the low-energy effective Lagrangian for $\varphi$ can be expressed as\begin{equation}\label{eq:L_eff}
{\cal L}_{\rm eff}[D_\rho \varphi] = P [(D_\rho \varphi D^\rho  \varphi)^{1/2}] \,,
\end{equation}
where  $P$ has now to be understood as a functional of the derivatives of the field $\varphi$ that has the same form as the pressure~\cite{Son:2002zn, 2008PhRvD..77j3014M}. 
In this context, we interpret the classical scalar field, $\bar \varphi$, as the potential field of the irrotational flow. Since the Lagrangian explicitly depends only on field derivatives, the classical equation of motion of  $\bar \varphi$ takes the form of the  hydrodynamic conservation law of the fluid current 
\be
 \partial_\nu (n v^\nu) = 0, 
\ee 
with 
\be n = \left.\frac{dP}{d \mu} \right\vert_{\mu =\bar \mu} \, \ee
the number density~\cite{Son:2002zn}, and 
$  v_\rho = {\bar \mu}^{-1} {D_\rho  \bar\varphi}\, $,
with $\bar \mu = (D_\rho  \bar \varphi D^\rho  \bar \varphi)^{1/2}$, the  fluid velocity, so that the velocity is properly normalized to one. 

From ~\eqref{eq:L_eff} it is possible to derive the effective field theory of the phonons moving in the background of the fluid \cite{Manuel:2007pz}. 
Our discussion applies to both  superfluid phonons and ideal-fluid hydrodynamics: indeed, in both cases the irrotational fluid motion can be described by the same scalar field Lagrangian, see for instance~\cite{SCHAKEL_1996}.
The  decomposition between the background fluid motion and the phonon oscillations can be obtained by a scale separation between the classical motion, described by $\bar \varphi$, and the long-wavelength  fluctuations associated to  the phonon field, described by $\phi$. 
Hence we write $\varphi (x) = \bar \varphi(x) + \phi(x)$ and expand the system low-energy action 
\begin{equation}
{\cal S}[\varphi] = \int d^4 x \, {\cal L}_{\rm eff}[\partial \varphi]\,
\end{equation}
around the classical solution
\begin{equation} \label{action2}
{\cal S}[\varphi] = {\cal S}[\bar \varphi] + \frac 12 \int d^4 x \,\frac{ \partial^2 {\cal L}_{\rm eff} }
{\partial(\partial_\mu \varphi) \partial(\partial_\nu \varphi)}
\Bigg \vert_{\bar \varphi}\partial_\mu \,\phi \partial_\nu \phi + \cdots \,.
\end{equation}
Considering  the Lagrangian in Eq.~\eqref{eq:L_eff},  one has  that 
\begin{equation}\label{eq:fmunu}
f^{\mu \nu} =  \frac{\partial{\cal L}_{\rm eff}}{\partial(\partial_\mu \varphi)
\partial(\partial_\nu \varphi)} \Bigg \vert_{\bar \varphi} = \frac{n}{\bar \mu}
\left \{ \eta^{\mu \nu} + \left(\frac {1}{c_s^2} - 1 \right) v^\mu v^\nu \right \} \,.
\end{equation}
This allows us to write  
\be
\label{phonon-action}
{\cal S}[\phi] = \frac 12 \int d^4 x \sqrt{- g } \, g^{\mu \nu} \partial_\mu \,\phi \partial_\nu \phi \,,
\ee
corresponding to the action of a boson in a  gravity background~\cite{Barcelo:2005fc} with the 
so-called acoustic  metric tensor~\cite{Bilic:1999sq,Visser:2010xv}
\begin{equation}
\label{phonon-metric}
g^{\mu \nu} = \Omega 
\left \{ \eta^{\mu\nu} + \left(\frac {1}{c_s^2} - 1 \right)  v^\mu  v^\nu \right \}\, ,
\end{equation}
with $\Omega$ a conformal factor. The inverse metric 
\be g_{\mu\nu}= \Omega^{-1} [\eta_{\mu \nu} + \left(c_s^2 - 1 \right)  v_\mu  v_\nu]\,, \ee
is obtained assuming that $v_\mu = \eta_{\mu\nu} v^\mu$.

For simplicity  we  take $\Omega=1$, although the presence of any conformal factor can be easily taken into account~\cite{2008PhRvD..77j3014M},   thus
$g \equiv \det{g_{\mu\nu}} = -c_s^2$.
If both the fluid and the sound speed are much smaller than the speed of light,
we recover the standard non-relativistic acoustic metric, see for instance \cite{Manuel:2007pz}. The acoustic horizon 
can be defined by the  condition that at the position $\bm x_H$ of the horizon, the $g_{tt}$ component of the metric vanishes, 
i.e. $c_s(\bm x_H) - v( \bm x_H)=0$. 

An AH, like a BH, behaves as a thermal reservoir~\cite{Brout:1995rd}, thus phonons are spontaneously emitted with a thermal distribution at the acoustic horizon, 
 $f(x,p)$, which in the long-wavelength limit is independent of the phonon-phonon interactions. Knowing $f(x,p)$, one can construct the phonon  energy-momentum tensor and entropy density, that using  a covariant formalism are respectively given by
\begin{eqnarray}
\label{eq:T}
T^{\alpha \beta}_{\rm ph}  & = & \int p^\alpha p^\beta   f(x,p) d {\cal P} \ , \\
\label{eq:S}
s^\alpha_{\rm ph}  & = &- \int   p^\alpha \left[ f \ln{f} - (1+f) \ln{(1+f)}\right] d {\cal P} \,,
\end{eqnarray}
with the covariant momentum measure~\cite{LINDQUIST1966487} 
\begin{equation}\label{eq:measure}
d {\cal P} = \sqrt{- g} 2H(p) \delta({ g}_{\mu \nu}p^\mu p^\nu) \frac{d^4p}{(2\pi)^{3}}\,,
\end{equation}
where $H(p) = 1$ if $p$ is future oriented for an observer moving with velocity $v_\mu$, and $0$ otherwise. These expressions turn to the standard ones in   Minkowski space-time. One  important aspect  is that the conserved quantities should include the metric~\cite{2008PhRvD..77j3014M}, in particular, for a collisionless fluid with no source term $(s^\alpha_{\rm ph})_{;\alpha} =0$
\begin{comment}
(a similar expression holds for the phonon density $n_{\rm ph}^\alpha$),
\end{comment}
(also $(n_{\rm ph}^\alpha)_{;\alpha}=0$ ), 
which is equivalent to
$(s^\alpha_{\rm ph})_{,\alpha} = - \Gamma^\mu_{\mu \nu} s_{\rm ph}^\nu = - {c_s}^{-1} (\partial_\nu c_s) s_{\rm ph}^\nu \, $. Thus an entropy (phonon) current, of purely geometrical origin, arises in the presence of gradients of $c_s$.

The stationary phonon distribution function, $f(x,p)$, should satisfy the covariant  Liouville equation~\cite{LINDQUIST1966487,stewart1969lecture}
\begin{equation}
\label{Boltzmman}
L[f] \equiv p^\alpha \frac{\partial f}{\partial x^\alpha} - \Gamma^\alpha_{\beta\gamma} p^\beta p^\gamma  \frac{\partial f}{\partial
p^\alpha}   = C[f]\,,
\end{equation}
corresponding to  the general relativistic version of the Boltzmann equation, where
the Christoffel symbols are obtained using Eq.~\eqref{eq:measure}. In the covariant kinetic theory,
$f(x,p) p^\mu n_\mu d \Sigma d {\cal P} $ represents the number of particles whose world lines intersect the
hypersurface element $n_\mu d \Sigma$ around $x$, having four-momenta in the range
 $(p, p +dp)$, and $n_\mu$ is a four light-like vector. 
The  stationary solutions of Eq.~\eqref{Boltzmman}
correspond to the condition $L[f]= 0$, \mm{meaning that we are neglecting any dissipative process} (see  Appendix~\ref{sec:append2} for a brief discussion of dissipation processes). If we assume that 
\begin{equation}
\label{eq:bose_dist}
f_{\rm}(x,p) = \frac{1}{\exp{(p^\mu  \beta_\mu)} - 1} \,,
\end{equation}
we obtain that the stationary configuration is characterized by the 4-vector 
$\beta_\mu$ that has to satisfy 
$ ( { \beta}_{\lambda, \rho} - { \beta}_\alpha \Gamma^\alpha_{\lambda \rho} ) \, p^\lambda p^\rho =0$,
which is equivalent to the Killing's equation~\cite{2008PhRvD..77j3014M}
\begin{equation}
\label{eq:Kill2}
{ \beta}_{\lambda; \rho} + { \beta}_{\rho;\lambda} = 0 \,,
\end{equation}
indicating how the equilibrium distribution function can be determined for any configuration with a given acoustic metric $g_{\mu \nu}$.

% - - - - -- - - - - 
\section{The spherical acoustic hole}
\label{sec:spherical}
% - - - - -- - - - - 

We shall now exploit the analogy with the BH thermodynamics to determine $\beta_\mu$. We  consider a spherically symmetric stationary flow with velocity
$\bm v = v \,\hat r\,$ along the radial direction 
$\hat r$ and $v(r)<0$, is the time independent velocity of the flow  falling towards the center. This is not a  physically realizable system,
due to a singularity at $r=0$, however it makes the analogy with standard BHs straightforward. More realistic configurations are considered later.
In spherical coordinates the line element reads
\begin{align}
ds^2 =& dt^2 (c_s^2-v^2) \gamma^2 + 2 \gamma^2(1-c_s^2) v dr dt\nonumber \\  &- [(1-c_s^2)\gamma^2 v^2 +1]dr^2 - r^2 d \Omega^2\,,
\end{align}
where $\gamma$ is the Lorentz factor, then
\begin{align}
g_{\mu\nu} = \left( \begin{array}{cccc} (c_s^2-v^2)\gamma^2 & (1-c_s^2) \gamma^2 v & 0 &0 \\ 
(1-c_s^2)\gamma^2 v &   (c_s^2-1)\gamma^2 v^2-1  &0  &0\\
0  &0 & -r^2  &0 \\
0  &0 & 0  & -r^2 \sin^2\theta  \end{array}\right)
\end{align}
and since  we assume stationary flow,  $g_{\mu\nu,t}=0$. 
We remark that at the acoustic horizon, phonons can only be emitted radially, otherwise they would have a radial velocity smaller than $c_s$ hence they cannot escape (see  Appendix~\ref{sec:append1} for more details).  Thus we define the $2 \times 2$ metric tensor
\begin{align}\label{eq:reduced_metric}
\tilde g_{\alpha\beta} = \left( \begin{array}{cc} (c_s^2-v^2)\gamma^2 & (1-c_s^2) \gamma^2 v  \\ 
(1-c_s^2)\gamma^2 v &   (c_s^2-1)\gamma^2 v^2-1\end{array} \right) \,,
\end{align}
where $\alpha,\beta = t, r$ so that $\tilde g = \det (\tilde g_{\alpha\beta}) = - c_s^2$ and
$g = \det (  g_{\mu\nu}) = \tilde g\, r^4 \sin^2\theta $.

It can be shown that $\beta_\mu = (\beta_t, \beta_r, 0,0 )$, 
meaning that  the system is effectively $1+1$ dimensional, thus phonons are only sensitive to the  $\tilde g_{\alpha\beta}$ metric. We shall assume $\beta_{\mu,t}=0$ as well.
The relevant Christoffel symbols are:
\begin{align}\label{eq:gammas}
\Gamma^{t}_{tt} &= - \frac{1}2 g^{tr}g_{tt,r}&\Gamma^{r}_{tt} = - \frac{1}2 g^{rr}g_{tt,r}\\
\Gamma^{t}_{rt} &=  \frac{1}2 g^{tt}g_{tt,r}  &\Gamma^{r}_{rt} =  \frac{1}2 g^{rt}g_{tt,r}\, ;
\end{align}
substituting these expressions in Eq.~\eqref{eq:Kill2} we obtain the expression for $\beta^\mu$.
For  $\lambda=\rho=t$,  Eq.~\eqref{eq:Kill2} yields
\be
g^{tr}\beta_t + g^{rr}\beta_r =0\,,
\ee
which implies $\beta^r =0 $. 
For $\lambda=r$ and $\rho=t$ we have
\be
0=\beta_{t,r} - 2 \Gamma^\alpha_{rt} \beta_\alpha =\beta_{t,r} - 2 \Gamma^t_{rt} \beta_t -  2 \Gamma^r_{rt} \beta_r\,,\ee
that using Eqs.~\eqref{eq:gammas} yields 
\be
\beta_{t,r} - g_{tt,r}(g^{tt}\beta_t + g^{tr}\beta_r)=0\,, 
\ee
while from $\lambda=r$ and $\rho=r$ we have
\be
\beta_{r,r} - g_{tr,r}(g^{tt}\beta_t + g^{tr}\beta_r)=0\,.\ee
From the above equations we readily have that
$\beta_t = \beta g_{tt}$ and $\beta_r = \beta g_{tr}$,
with $\beta$ a  constant, implying that $\beta^\mu = (\beta,\bm 0)$. 
i.e. the stationary condition implies that the phonon distribution function is completely determined by only one parameter $\beta=1/T$, that has no angular or radial dependence.

To fix $T$ we now need to link the phonon distribution function to  the acoustic  horizon.  Since the kinetic equations describe on-mass shell particles, we first look at 
 the phonon  dispersion law. Phonons follow null geodesics of the acoustic metric,  $g^{\mu\nu}p_{\mu}p_\nu =0$: writing $p_\mu=(E, -p_r,0,0)$  (see Appendix \ref{sec:append1}) we obtain 
\be
g^{tt} E^2 - 2 g^{tr} E p_r + g^{rr} p^2_r =0\,,
\ee
with solution $E_\pm = K_\pm p_r  $
and $K_\pm = (v \pm  c_s)/(1\pm c_s v)$
the effective phonon velocity. 
In the considered system $v<0$ 
\be\label{eq:Kpm}
K_\pm = \begin{cases}
\displaystyle{\frac{c_s - |v|}{1- c_s |v|} } & \phantom{for}
\\[1.ex]
\displaystyle{ \frac{-c_s - |v|}{1+ c_s |v|} } & \phantom{for }
\end{cases},
\ee
meaning that outside the acoustic horizon, where $|v|<c_s$,  the mode with $p_r>0$ has positive energy.   Inside the horizon, both modes propagating towards the center with $p_r<0$,  have positive energy, \mc{though} only one  mode is relevant for our discussion.
This is more evident in the fluid rest frame ($v=0$) where the  phonons have dispersion law  $E_\pm = \pm c_s p_r$, therefore there exist two radially emitted modes with positive energy, one propagating outwards and one inwards.
The sum of the energies of the emitted phonons is not zero, therefore phonons carry away  energy from the acoustic horizon.

We now proceed to extend the entropy area law of BHs to AHs, allowing us to relate $\beta=1/T$ to the Hawking temperature.
As for BHs, we assume that it is possible to associate to the acoustic horizon the entropy \mm{
\be
S_H =\kappa \frac{A}{L_c^2}  =  \frac{4 \kappa  \pi r_H^2}{L_c^2} \,,\ee
}where $L_c$ is the phonon cutoff length scale playing the same role of the Planck length scale in General Relativity~\cite{Jacobson:1991gr}, $\kappa$ is a number to be fixed, while $A$ and $r_H$ are the area and radius of the acoustic horizon, respectively.   
We note that this result can in principle be formally derived along the same reasoning lines used to determine the information loss~\cite{Bekenstein:1973ur} or the entanglement of quantum fluctuations on the two sides of the horizon~\cite{Jacobson:1994iw}.  If the AH emits sound waves by a radial fluctuation, then in the laboratory frame its entropy variation is 
\be
d S_H = 8  \pi \kappa \frac{r_H }{L_c^2} d r_H \,,
\ee
where  $dr_H$ is a small variation of the horizon radius.  The corresponding variation of the phonon entropy in the volume around $r_H$ is instead 
\be\label{eq:dSph}
dS_{\rm ph} = 4  \pi  r_H^2   \,  d_g \,  \tilde s_{\rm ph} d r_H\,,
\ee 
where $ \tilde s_{\rm ph}$ is the conserved entropy  density of phonons. The number $d_g=3 \times 2$ takes into account the contributions of the three phonon quasiparticles  and of the modes propagating  outside and inside the AH. As discussed in \cite{Macher:2009nz, Macher:2009tw} for a superfluid (see \cite{Leutwyler:1996er,SCHAKEL_1996} for ideal fluids), when counting the number of modes in our three-dimensional superfluid geometry, we need to account for 3 spatial degrees of freedom and for the existence of Bogoliubov $v$ and $u$ modes, propagating inside  and outside the AH, respectively.  This mode counting is independent of the details of phonon propagation.
% which we will come back to below. 

In our base hypothesis, phonons are produced only by geometry. Thus, upon equating $\displaystyle{ d S_H = dS_{\rm ph}}$ we obtain 
\be\label{eq:stildeph}
\tilde s_{\rm ph} = \frac{\kappa}{3 r_H L_c^2 }\,,
\ee
which relates the phonon entropy density to a geometric property of the system. In evaluating  
$\tilde s_{\rm ph}$, we have now to consider that the horizon can only emit phonons radially. As shown in Appendix \ref{sec:append1}, this fact, together with the presence of the scale $L_c$, affects the momentum measure which should be cast as (see Eq.~\eqref{eq:A6})
\mm{
\begin{equation}
d {\cal P} = c_s 2H(p) \delta({ \tilde g}_{\mu \nu}p^\mu p^\nu) \frac{d p^0 dp^r}{2\pi L_c^2}\,,
\end{equation}
where ${ \tilde g}_{\mu \nu}$ is the  two-dimensional metric, see  Eq.~\eqref{eq:reduced_metric}}. 

Using the results derived in Appendix \ref{sec:append2},
we are now ready to express the phonon entropy density Eq.~\eqref{eq:S} close to the  horizon, obtaining that it is directly proportional to the temperature:
\be\label{eq:entropy2d}
\tilde s^0_{\rm ph} = \tilde  s_{\rm ph} = \frac{\pi T}{6 K_+ L_c^2}\,,
\ee 
\mc{with $K_+$ given by Eq.~\eqref{eq:Kpm}}. Substituting this expression in Eq.~\eqref{eq:stildeph}, \mm{the cutoff length scale simplifies}  \mc{and}
the AH temperature \mc{turns out to be} \mm{
\be\label{eq:temp}
T = \frac{2 \kappa K_+}{\pi r_H}\,.
\ee
}After shifting  the radial  coordinate so that the acoustic horizon corresponds to $r_H=0$, and
 assuming that close to the acoustic horizon $v+c_s= C r$,
$C= (v +c_s)'_H$, Eq.~\eqref{eq:temp}  yields
\mm{\be\label{eq:TH}
T = \frac{2 \kappa}{\pi} \left.\left(\frac{ c_s-|v|}{1-c_s |v| }\right)'\ \right\vert_H\,,
\ee
}\mc{in fact} equivalent to Eq.~\eqref{eq:hawking} in the nonrelativistic limit for $\kappa=1/4$. \dg{Remarkably this is the same value generally adopted for BHs (see {\it e.g.} \cite{Bekenstein:1973ur})}.   

\mc{Notice that} the same result can be obtained in a similar way by equating the energy loss of the AH and  the  energy gain of the phonon gas.  In this case, we assume that the spontaneous phonon emission is associated to \mc{the energy variation of an AH}
\be
\label{eq:energyH}
d E_H \propto \frac{1}2  \frac{d r_H}{L_c^2} \,,
\ee
which is \mc{indeed} the analogue of the BH energy variation by spontaneous photon emission.  The corresponding  energy variation of the phonon gas is 
\be\label{eq:dEph}
d E_\text{ph} = 4 \pi r_H^2 d_g   (\tilde T_{\rm ph})^{0}_{\phantom{0}\,0} d r_H\,,
\ee
where $d_g$ is the degeneracy introduced above and  $  (\tilde T_{\rm ph})^{0}_{\phantom{0}\,0} = \sqrt{-\tilde g} (T_{\rm ph})^{0}_{\phantom{0}\,0}$. Equating $d E_H=d E_\text{ph}$ and  using \eqref{eq:T} we obtain again the expression of the Hawking temperature in ~\eqref{eq:TH}. 

These results indicate that spherical AHs, as ordinary BHs,  can be viewed as a thermodynamic system at fixed temperature spontaneously emitting massless particles carrying energy and entropy. In AHs, the emission can be viewed as  triggered  by  fluctuations of the acoustic horizon that produces radially propagating sound waves.
Since the spherical acoustic horizon is exactly isothermal, to make the analogy with BHs  more stringent, one should actually consider a BH that accretes matter by a rate exactly equivalent to the energy loss by photon emission. This difference is relevant for light BHs, possibly produced at LHC~\cite{Eardley:2002re,Yoshino:2002tx,Yoshino:2005hi}, however in general the BH mass is much larger than the radiated energy and the process can be  approximated as isothermal~\cite{Misner:1974qy, Shapiro:1983du}. 

% - - - - - - - - - 
\section{More realistic geometries}
\label{sec:realistic}
% - - - - - - - - - 

Although we have derived the above relations  using a spherically symmetric system, our results hold \mc{as well} for different geometries. 
\begin{figure}[ht!]
\includegraphics[width=0.45\textwidth]{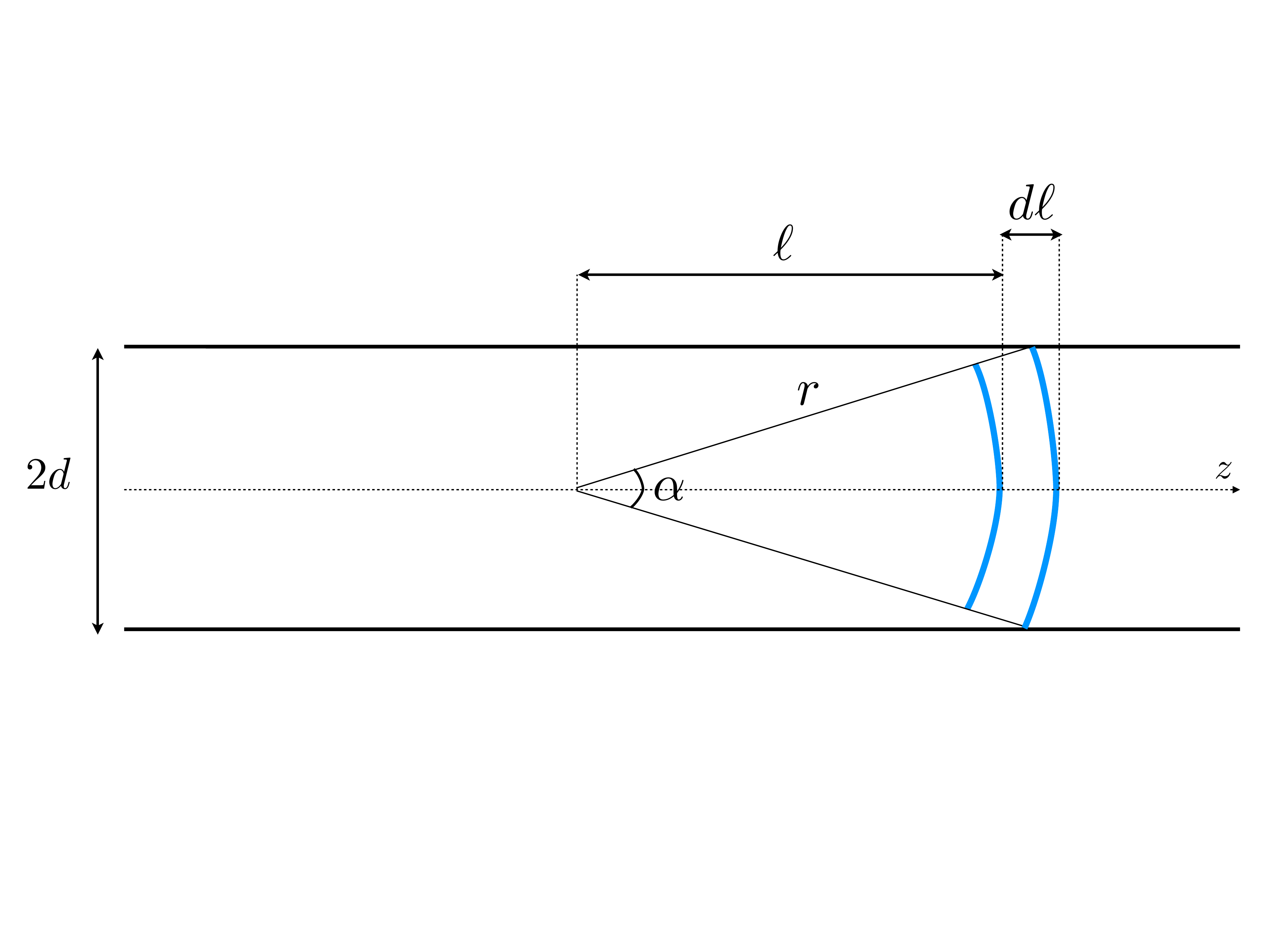}\vspace*{-1.cm}
\caption{Schematic representation of the section of a cylindrical tube with transonic flow. The blue curves represent the acoustic horizon during phonon emission, \mc{causing the horizon's radial fluctuation with its slight reduction between the blue lines, followed by a corresponding increase of the phonon density in the volume within the two.}}
\label{fig:tube}
\end{figure}
As an example, in Fig.~\ref{fig:tube} we schematically show the variation of the acoustic horizon for a flow in a cylindrical tube of diameter $2 d$ and axis in the $z$-direction. 
For a viscous fluid, the effective area of the acoustic horizon can be approximated by $A = c_s \alpha r^2$, where $r=\sqrt{d^2+\ell^2}$;
for an unviscous flow, the acoustic horizon becomes flat
 corresponding to the $\ell \to 0$ limit. Assuming that the phonon emission happens by a horizon radial fluctuation and that it is possible to associate to the acoustic horizon an entropy as in BHs,  the entropy change of the fluid is
\be
\label{eq:entropyH}
d S_H = \frac{2 \kappa \alpha c_s \ell d\ell}{ L_c^2}\,, 
\ee
while the associated change  in the phonon entropy is
\be
dS_\text{ph}=dV  \tilde s_\text{ph}= c_s \alpha \ell^2 d \ell \tilde s_\text{ph}\,.
\ee
Equating $d S_H  = dS_\text{ph} $ and using  the phonon entropy density in Eq.~\eqref{eq:entropy2d}, 
we obtain $\displaystyle {T = \frac{ 2 \kappa K_+}{ \pi d_g  \ell}}$ and
\be\label{eq:temp_tube}
 \lim_{\ell \to 0} T =    \frac{2 \kappa}{\pi} \left. \frac{\partial}{\partial z} \left( \frac{ c_s-|v|}{1-c_s |v| } \right) \right\vert_H\,,
\ee
that  is the  same expression of the Hawking temperature reported in Eq.~\eqref{eq:TH}. 
As for the spherical case, we obtain the same expression of the Hawking temperature when equating the  energy loss of the AH and the energy gain of the phonon gas. The present derivation can be straightforwardly extended  to reduced space dimensions, {\it e.g.} in 2D, by appropriately counting the corresponding number of phonon modes.

% - - - - - - - - -
\section{Conclusions}
\label{sec:conclusions}
% - - - - - - - - -

We have shown how for AHs the Hawking temperature comes with a conceptually simple covariant kinetic theory approach. Under the hypothesis that the BH and AH entropies have the analogous physical origin, we associate to the latter the entropy  $A/4$ in terms of the horizon surface area $A$. Then,  we determine the Hawking temperature by equating  the entropy  loss  of the fluid with  the phonon entropy gain. The same result can be obtained associating to an AH a a mass equal to half the Schwarzschild radius and equating the energy loss of the fluid with the energy gain of the phonon gas. These results hold if the horizon fluctuates radially, which is appropriate for sonic wave emission. 

Our findings elucidate the relationship between the fluid fluctuations and the phonon (Hawking) emission
gas variations, further clarifying the geometrical nature of the process.
Remarkably, our results do not depend on  the phonon emission mechanism and are truly shaped only by the geometrical properties of the acoustic horizon, as clearly emerges from our covariant description. However, the proposed method works only in a region sufficiently close to the acoustic horizon, where the system is effectively $1+1$ dimensional, and for a  fluid temperature much lower than  $T_H$.  In this case, it is possible to neglect phonon-phonon interactions,   meaning that the spontaneously emitted phonon gas is in a out-of-equilibrium thermodynamic state ~\cite{prigogine2017non}. 

Interestingly, we are also able to clarify an important aspect of BH physics: the spontaneous emission of photons at the event horizon should result in energy and entropy gains of the photon gas equal to the mass and entropy losses of the BH. In this case though, one  complication arises because of the long-range behavior of the gravitational interaction, which may not allow to separate the photon energy density and entropy from the BH ones~\cite{Brout:1995rd}.

The physics described here can be probed in currently available experimental setups based on extremely accurate control of table-top quantum technologies, such as ultracold gases platforms. In particular, we envisage that the 2D, disk-shaped geometry setup experimented in~\cite{Chin2017,Chin2019} can be especially promising in probing the behavior we describe. In the experiment, a coherent Hawking radiation is simulated in a 2D Bose-Einstein condensate  after modulating the scattering length driving the atomic interactions, and thus the speed of sound, by a variable amount of time. The measured probability distribution of the matter-wave emission is observed to be thermal.
Adapting Eqs.~\eqref{eq:dSph} and \eqref{eq:dEph} to the desired geometry one can compute the rate of entropy as well as of energy emitted by the AH. These quantities could be measured by experimental strategy adopted in ~\cite{Zwierlien:2012} or via measurements of dynamical correlation functions~\cite{Steinhauer2}. In this measurement one should  take into account the bulk temperature of the system, here neglected, as well as the transport properties of the superfluid. 
Quite generally, the theoretical framework here developed opens to new formulations of the AH physics with unexplored perspective, shining a different light on the treatment dynamical and dissipative effects.
 
\begin{acknowledgments} We acknowledge useful discussions
with  Stefano Liberati and Iacopo Carusotto.
\end{acknowledgments}

\appendix
% - - - - - - - - -
\section{Dimensional reduction close to the acoustic horizon}
\label{sec:append1}
% - - - - - - - - -

\begin{figure}[h!]
\includegraphics[width=0.45\textwidth]{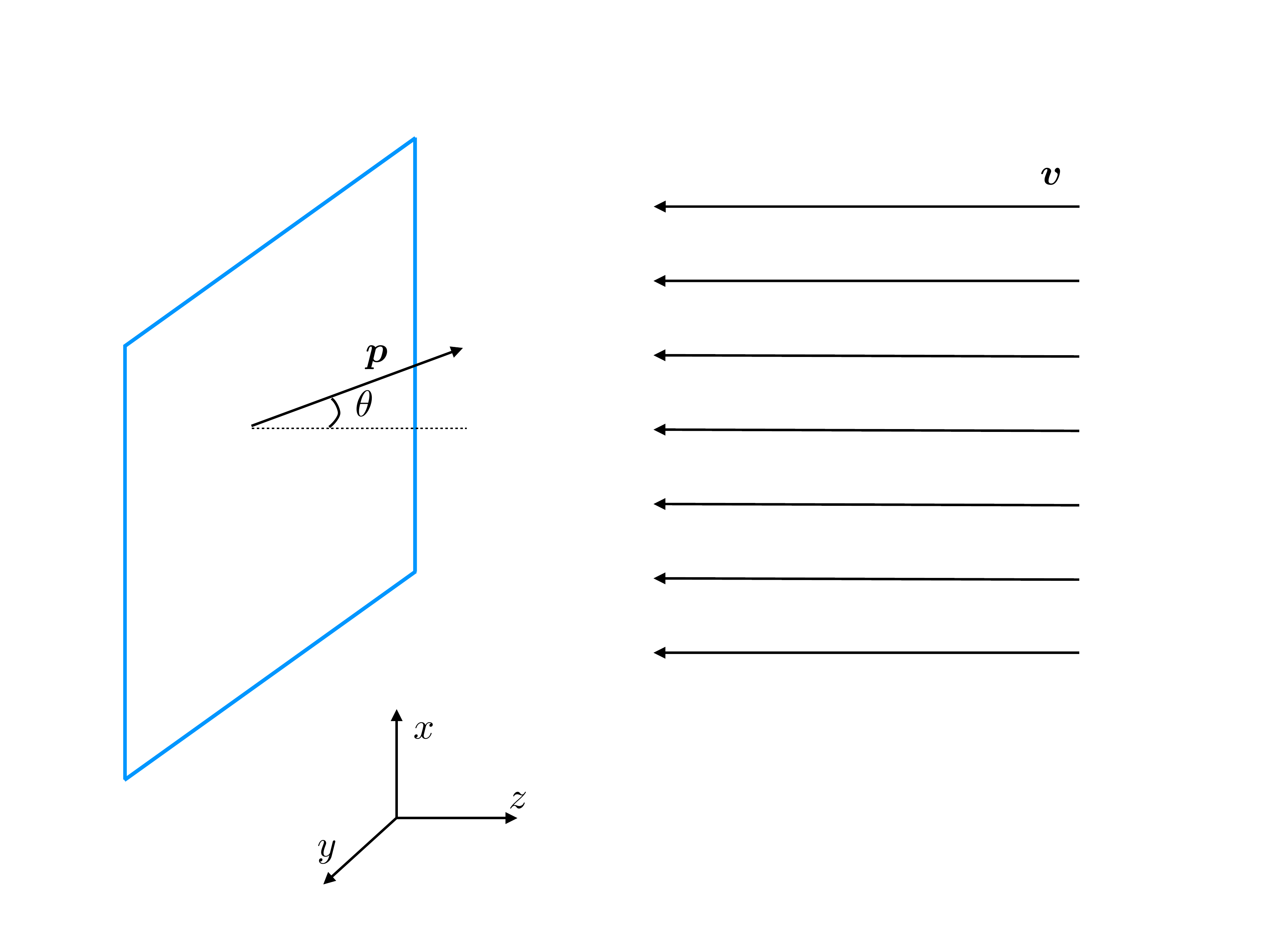}\vspace*{-0.cm}
\caption{Schematic representation of the emission of a phonon close to the acoustic horizon. The blue rectangle corresponds to the local surface of the acoustic horizon orthogonal to  the fluid velocity $\bm v$, which is along the $z$ axis. The black line corresponds to the emitted phonon with momentum ${\bm p}$ at a certain angle $\theta$ with respect to the normal to the surface. Only phonons emitted with vanishing $\bm p_x$ and $\bm p_y$ components  can escape from the acoustic  horizon.  }
\label{fig:tube2}
\end{figure}

The acoustic horizon is the region where the fluid velocity equals the speed of sound \cite{Unruh:1980cg,Brout:1995rd,  Barcelo:2005fc}. Since phonons propagate at the speed of sound, the only phonons that can escape from the acoustic horizon are those having velocity with vanishing azimuthal angle with respect to the fluid motion, corresponding to   vanishing values of $\theta$ in Fig.~\ref{fig:tube2}, where  the horizon is pictorially represented by the blue  contour. In  fluids, the phonon momentum and velocity are parallel, thus  only those  phonons emitted with  vanishing  $p_x$ and $p_y$ components can escape. We can interpret this result assuming that the distribution function of phonons close to the horizon of Fig.~\ref{fig:tube2} is given by
\be
f_\text{ph} = f\cdot  f_t(p_x,p_y)\,,
\ee
where $f$ is the standard Bose-Einstein distribution (see Eq. \eqref{eq:bose_dist}), while the  $f_t$ distribution should in principle be derived by the microscopic physics. However, sufficiently close to the acoustic horizon  $f_t$  can  only depend on the cutoff length scale $L_c$, which is the  only relevant length scale. For this reason  we assume that sufficiently close to the horizon of Fig.~\ref{fig:tube2} the transverse distribution function is given by
\be
 f_t(p_x,p_y)= \frac{2\pi}{L_c}\delta(p_x)  \frac{2\pi}{L_c}\delta(p_y)\,.
\ee
This expression can be generalized as 
\be
 f_t(p_t) = \left(\frac{2\pi}{L_c}\right)^2 \delta(p_t^2)\,
\ee 
close to any acoustic horizon, where $p_t$ is the transverse momentum with respect to the local orientation of $\bm v$ at the horizon, while we will indicate with $p^r$ the momentum parallel to $\bm v$.
Instead of the Dirac delta function, one may consider a  Gaussian distribution  or any  representation of the Dirac delta function, the results would be qualitatively the same. 
In any case, this would be still consistent with our kinetic theory approach (see Eq.~\eqref{Boltzmman}).
When evaluating  the energy-momentum tensor (see for instance \cite{2008PhRvD..77j3014M})
\be
\label{eq:Tn}
T^{\alpha \beta}_{\rm ph}  =  \int p^\alpha p^\beta   f_\text{ph}(p) d {\cal P} \ , \ee
the momentum  measure \cite{LINDQUIST1966487,stewart1969lecture}
\be\label{eq:measureP}
d {\cal P} = \sqrt{- g} 2H(p) \delta({  g}_{\mu \nu}p^\mu p^\nu) \frac{d p^0 dp^r dp_t^2}{(2\pi)^3}\,,
\ee
can  be simplified by  integrating out  the transverse momentum. Taking into account  that $\sqrt{- g}=c_s$, the resulting  integration measure is
 \be
d {\cal P} = c_s 2H(p) \delta({ \tilde g}_{\mu \nu}p^\mu p^\nu) \frac{d p^0 dp^r}{(2\pi) L_c^2}\,,
\label{eq:A6}
\ee
where    ${ \tilde g}_{\mu \nu}$ is the  two-dimensional metric. This expression can be rewritten as
\be\label{eq:measure_simp}
d {\cal P} =  c_s \frac{\delta(p_0-E_+)}{ p^0} \frac{d p^0 dp^r}{(2\pi)L_c^2}\,,
\ee
where $E_+$ is the dispersion law of phonons. The latter  can be obtained from $g^{\mu\nu}p_\mu p_\nu=0$ and has the general expression
\be
\label{eq:disprel}
E_\pm= \frac{v p_r(1-c_s^2) \pm c_s \gamma^{-2} \sqrt{p_t^2 \gamma^2(1-c_s^2 v^2)+p_r^2}}{1-c_s^2 v^2}\,,
\ee
which for vanishing transverse momenta gives
\be\label{eq:E+n}
E_\pm=\frac{v \pm c_s}{1\pm v c_s} p_r  = K_\pm p_r \,,
\ee
with the $+$ sign corresponding to the positive energy states relevant for the kinetic equations.

% - - - - - - - - -
\section{Evaluation of the thermodynamic quantities}%---
\label{sec:append2}
% - - - - - - - - -

From the expression of the energy-momentum tensor in Eq.~\eqref{eq:Tn} and using the integration measure in Eq.~\eqref{eq:measure_simp} we can extract the energy density
\be\label{eq:endensity}
\tilde \epsilon_\text{ph} = \sqrt{-g} T^0_{\phantom{0}0}=\int E_+ f \frac{dp_r}{2 \pi L_c^2}\,,
\ee
where 
\be\label{eq:dist_onshell}
f=\frac{1}{e^{K_+ p_r/T} -1}\,,
\ee
is the phonon distribution function. Similarly,  the pressure can be evaluated by using
\be
\sqrt{-g} T^i_{\phantom{i}j} = \delta^{ir} \delta_{jr} \tilde  P_\text{ph}\,,
\ee
showing that the pressure is radially oriented:  it is locally orthogonal to the acoustic horizon. 
We can  derive the entropy density by  the thermodynamic relation
\be
\tilde \epsilon_\text{ph} + \tilde  P_\text{ph} = T \tilde  s_\text{ph}\,,
\ee
which  holds because phonons are emitted at the temperature, $T$, and they  have vanishing chemical potential. Since  $\tilde P_\text{ph}=\tilde \epsilon_\text{ph}$,  we obtain that 
$$
T \tilde s_\text{ph} =  2 \tilde \epsilon_\text{ph}\,,
$$
and upon changing the integration variable  to $x= K_+ p/T$,  Eq.~\eqref{eq:endensity}  yields
\be
\tilde \epsilon_\text{ph} = \frac{T^2}{2 \pi L_c^2 K_+}  \int_0^{\infty}  x f(x) dx \,.
\ee
Since
\be
\int_0^{\infty}  x f(x) dx = \frac{\pi^2}{6}\,,
\ee
it follows that 
\be
\label{eq:ced}
\tilde s_\text{ph}  = \frac{\pi T}{ 6 L_c^2 K_+}  \,.
\ee

Alternatively, the entropy density can be evaluated using the expression
\be
s_{\rm ph}  = - c_s \int   \left[ f \ln{f} - (1+f) \ln{(1+f)}\right]  \delta(p_0-E_+) \frac{d p^0 dp^r}{(2\pi) L_c^2} \,,
\ee
and upon lowering the indices in the integration measure,  which amounts to divide by $c_s^2$, and integrating over $p_0$ we obtain
\be
s_{\rm ph}  = - \frac{1}{c_s} \int   \left[ f \ln{f} - (1+f) \ln{(1+f)}\right]   \frac{dp_r}{(2\pi) L_c^2} \,,
\ee
where the distribution function is in Eq.~\eqref{eq:dist_onshell}. 
Upon changing the integration variable to $x= K_+ p_r/T$, we get
\be
s_{\rm ph}  = - \frac{T}{ (2 \pi) L_c^2 K_+ c_s} \int   \left[ f \ln{f} - (1+f) \ln{(1+f)}\right]  d x  \,,
\ee
where the integral can now be analytically evaluated to
\be
\int   \left[ f \ln{f} - (1+f) \ln{(1+f)}\right]  d x  = -\frac{\pi^2}{3}\,,
\ee
meaning that the entropy density $\tilde s_\text{ph}=\sqrt{-g} s_\text{ph}$, takes the expression in Eq.~\eqref{eq:ced}.
Noticeably, for $r \rightarrow r_H$, the factor
$K_+$ introduces a divergence in the  energy and enetropy densities, meaning that we are approaching the phonon source.
However, the covariant entropy and energy density in  Eqs.~\eqref{eq:endensity} and \eqref{eq:ced}  can be regularized by the short distance cutoff $L_c$.
This is evident from \eqref{eq:disprel}, accounting for the finite value of the transverse momentum $p_t \sim 1/L_c^2$,
hence removing the divergence that appears in the energy and entropy densities due to the vanishing of $K_+$ at the acoustic horizon. 
The non-vanishing transverse momentum can be related to dissipative effects, which transfer  momentum  in the orthogonal direction with respect to the flow. This shows that dissipation can effectively regularize the  behaviour close to the horizon. 
We shall come back on this issue in a separate work.

\bibstyle{ieeetr}
\bibliography{Phonon_PRD}

\end{document}